\documentclass[lettersize,journal]{IEEEtran}
\usepackage{amsmath,amsfonts}
\usepackage{algorithmic}
\usepackage{array}
\usepackage{textcomp}
\usepackage{stfloats}
\usepackage{url}
\usepackage{verbatim}
\usepackage{graphicx}
\usepackage{cite}

\usepackage{multirow}
\newcolumntype{P}[1]{>{\centering\arraybackslash}p{#1}}
\newcolumntype{M}[1]{>{\centering\arraybackslash}m{#1}}
\usepackage{algorithm}

\usepackage{colortbl}
\definecolor{Gray}{gray}{0.85}
\definecolor{LightCyan}{rgb}{0.88,1,1}

\def\BibTeX{{\rm B\kern-.05em{\sc i\kern-.025em b}\kern-.08em
    T\kern-.1667em\lower.7ex\hbox{E}\kern-.125emX}}
\usepackage{balance}
\usepackage{subcaption}

\begin{document}

\title{Noisy Neonatal Chest Sound Separation for High-Quality Heart and Lung Sounds}

\author{\uppercase{E. Grooby}, \IEEEmembership{Student Member, IEEE},
\uppercase{C. Sitaula},
\uppercase{D. Fattahi},
\uppercase{R. Sameni}, \IEEEmembership{Senior Member, IEEE},
\uppercase{K. Tan},
\uppercase{L. Zhou},
\uppercase{A. King},
\uppercase{A. Ramanathan},
\uppercase{A. Malhotra},
\uppercase{G.A. Dumont}, \IEEEmembership{Life Fellow, IEEE}, \uppercase{and}
\uppercase{F. Marzbanrad}, \IEEEmembership{Senior Member, IEEE}
\thanks{E. Grooby  acknowledges the support of the MIME-Monash Partners-CSIRO sponsored PhD research support program and Research Training Program (RTP). A. Malhotra research is supported by the Kathleen Tinsley Trust and a Cerebral Palsy Alliance Research Grant. The study is supported by Monash Institute of Medical Engineering (MIME).}
\thanks{E. Grooby, C. Sitaula and F. Marzbanrad are with the Department of Electrical and Computer Systems Engineering, Monash University, Melbourne, VIC, Australia.}
\thanks{D. Fattahi is with Shiraz University, Shiraz, Iran.}
\thanks{R. Sameni is with the Department of Biomedical Informatics, Emory University, Atlanta, GA, USA}
\thanks{K. Tan, L. Zhou, A. King, A. Ramanathan and A. Malhotra are with Monash Newborn, Monash Children’s Hospital and Department of Paediatrics, Monash University, Melbourne, Australia.}
\thanks{G. Dumont is with the Department of Electrical and Computer Engineering, University British Columbia, Vancouver, BC, Canada and with the BC Children's Hospital Research Institute, Vancouver, BC, Canada.}
\thanks{email: ethan.grooby@monash.edu}}

\maketitle

\begin{abstract}
Stethoscope-recorded chest sounds provide the opportunity for remote cardio-respiratory health monitoring of neonates. However, reliable monitoring requires high-quality heart and lung sounds. This paper presents novel Non-negative Matrix Factorisation (NMF) and Non-negative Matrix Co-Factorisation (NMCF) methods for neonatal chest sound separation. To assess these methods and compare with existing single-source separation methods, an artificial mixture dataset was generated comprising of heart and lung and noise sounds. Signal-to-noise ratios were then calculated for these artificial mixtures. These methods were also tested on real-world noisy neonatal chest sounds and assessed based on vital sign estimation error and a signal quality score of 1-5 developed in our previous works. Additionally, the computational cost of all methods was assessed to determine the applicability for real-time processing. Overall, both the proposed NMF and NMCF methods outperform the next best existing method by 2.7\,dB to 11.6\,dB for the artificial dataset and 0.40 to 1.12 signal quality improvement for the real-world dataset. The median processing time for the sound separation of a 10\,s recording was found to be 28.3\,s for NMCF and 342\,ms for NMF. Because of stable and robust performance, we believe that our proposed methods are useful to denoise neonatal heart and lung sound in a real-world environment. Codes for proposed and existing methods can be found at: \url{https://github.com/egrooby-monash/Heart-and-Lung-Sound-Separation}.
% For vital sign estimation, further work is required for accurate estimation for real-world chest sounds obtained from neonates on respiratory support.

\end{abstract}

% Blind single-source separation

\begin{table*}[tb]
    \begin{center}
\caption{Existing Blind Single-Source Separation Methods.}
    \begin{tabular}{ |M{3cm}|M{14.2cm}|}
    \hline \shortstack[c]{\textbf{Method}\\\textbf{(letter reference*)}}
    & \textbf{Description}
    \\
    \hline
    Adaptive Fourier Decomposition (c)
    & Based on energy distribution, adaptive Fourier decomposition is used to clean identified heart sound S1 and S2 peaks \cite{wang2015adaptive,springer2015logistic,grooby2020neonatal}. Top 5, 10, 15, 20 and 25 decomposition levels are assigned as heart sounds was tested. 
    %Decomposes signal based on energy distribution. Heart segmentation is performed to identify S1 and S2 peaks \cite{springer2015logistic,grooby2020neonatal}. Adaptive Fourier decomposition is performed on each peak with top x decomposition levels assigned heart and remaining as lung sounds \cite{wang2015adaptive}. Top 5, 10, 15, 20 and 25 levels assigned as heart sounds was tested.
    \\
    \hline
    Adaptive Line Enhancement (d)
    & Applies adaptive filtering on the original and time-delayed recording to extract semi-periodic components (heart sounds) \cite{tsalaile2007separation}. Time delay values of 1, 10, 40, 100, 200 and 400 samples are tested. 
    %Time-delayed and original mixture recording inputted into the adaptive filter. The adaptive filter extracts a semi-periodic component (heart sound) as output with the error signal being wide-band noise (lung and noise sounds) \cite{tsalaile2007separation,basak2010phonocardiogram}. Time delay values of 1, 10, 40, 100, 200 and 400 samples are tested. 
    \\
    \hline
    Empirical Mode Decomposition (e)
    & Decomposes signal into a sum of oscillatory functions called IMF. For each MF, S1 and S2 peaks are identified and filtered to obtain heart and lung components \cite{mondal2011reduction}. Empirical mode decomposition, ensemble empirical mode decomposition and complete ensemble empirical mode decomposition with adaptive noise are tested.
    %Empirical mode decomposition decomposes mixtures into a sum of oscillatory functions called IMFs. For each IMF, S1 and S2 peak detection is performed. Heart included segments are highpass filtered to obtain just the lung components and summed together with heart-free components to get lung sounds \cite{mondal2011reduction,lin2013lung}. Empirical mode decomposition, ensemble empirical mode decomposition and complete ensemble empirical mode decomposition with adaptive noise are tested.
    \\
    \hline
    Filtering (f)
    & 4\textsuperscript{th}-order Butterworth bandpass filter with passband frequencies 50-250\,Hz and 200-1000\,Hz, was used to obtain heart and lung sounds, respectively \cite{grooby2020neonatal,grooby2021real}.
    %4\textsuperscript{th}-order Butterworth bandpass filter with passband frequencies 50-250\,Hz and 200-1000\,Hz, was used to obtain heart and lung sounds, respectively. These frequency bands were determined based on past works \cite{grooby2020neonatal,grooby2021real}.
    \\
    \hline
    Interpolation (g)
    & Identified heart sound S1 and S2 peaks are removed in time-frequency domain either by 20-300\,Hz bandstop filter or complete elimination of them sections \cite{pourazad2006heart,springer2015logistic,grooby2020neonatal}. Interpolation is then performed to recover the lung sounds in the removed segments \cite{pourazad2006heart}.
    %Heart segmentation is performed to identify S1 and S2 peaks \cite{springer2015logistic,grooby2020neonatal}. In the time-frequency domain, at the sites of the S1 and S2 peaks, either a 20-300\,Hz bandstop filter or complete elimination of them sections is performed. Linear 2D interpolation is then performed to recover the lung sounds in the removed segments \cite{pourazad2006heart}. 
    \\
    \hline
    Modulation Filtering (h)
    & Involves bandpass and bandstop filtering of temporal trajectories of short-term spectral components to obtain heart and lung sounds respectively \cite{falk2008modulation}. Filter ranges tested were 1-20\,Hz, 2-20\,Hz, 3-20\,Hz, 4-20\,Hz, 5-20\,Hz and 6-20\,Hz. 
    %Modulation filtering refers to filtering the temporal trajectories of short-term spectral components to obtain clear heart and lung sounds. In the time-frequency domain, a bandpass filter is used to obtain heart sound components and a bandstop for lung components \cite{falk2008modulation}. Filter ranges tested were 1-20\,Hz, 2-20\,Hz, 3-20\,Hz, 4-20\,Hz, 5-20\,Hz and 6-20\,Hz. 
    % \cite{falk2008modulation,mardiyanto2013study}
    \\
    \hline
    NMF Clustering 1 and 2 (i and j)
    & Both methods blindly decompose the mixture into numerous sub-components and then cluster all components into either heart or lung based on spectral or temporal criteria \cite{grooby2021new,shah2014blind,canadas2017non}. 
    %Two existing methods were reviewed and adapted in our past work \cite{grooby2021new}. Briefly, both methods blindly decompose the mixture into numerous sub-components and then cluster all components into either heart or lung based on spectral or temporal criteria \cite{shah2014blind,canadas2017non}.
    \\
    \hline
    Recursive Least Squares Adaptive Filtering (k)
    & Identified heart sound S1 and S2 peaks used to create reference heart sound \cite{gnitecki2003recursive,springer2015logistic,grooby2020neonatal}. The recursive least squares filter uses original recording and the reference heart sound to obtain clean heart and lung sounds \cite{gnitecki2003recursive}.
    %Heart segmentation is performed and used to construct reference heart sound which equals 20-300\,Hz bandpass filtered mixture recording at the S1 and S2 locations and zero elsewhere \cite{gnitecki2003recursive,springer2015logistic,grooby2020neonatal}. Original mixture recording and reference heart sound are inputted into the recursive least squares adaptive filter. The output of the adaptive filter is heart sound and the error signal corresponds to lung sounds \cite{gnitecki2003recursive,lee2010single}. 
    \\
    \hline
    Singular Spectrum Analysis (l)
    & Decomposes signal into principal components using singular value decomposition \cite{ghaderi2011localizing}. Principal components associated with top eigenvector pairs with the strongest frequency component less than 250\,Hz are assigned as heart sounds, and the remaining are assigned as lung sound \cite{ghaderi2011localizing,mondal2017noise}. 
    %Mixture recording is decomposed into principal components using singular value decomposition \cite{ghaderi2011localizing}. Principal components with small eigenvalues are assumed to be noise and removed. Principal components associated with top eigenvector pairs with the strongest frequency component less than 250\,Hz are assigned as heart sounds, and the remaining are assigned as lung sound \cite{ghaderi2011localizing,mondal2017noise}. 
    \\
    \hline 
    Wavelet Transform based Stationary Non-Stationary Filter (m)
    & Wavelet thresholding is used to separate station sounds (lung sounds) from non-stationary sounds (heart sounds using an adaptive threshold with adjusting multiplicative factor x \cite{hadjileontiadis1998wavelet,hossain2003overview}. Adjusting multiplicative factors of 2, 2.5, 2.7, 3, 3.5 and 4 were tested. 
    %S1 and S2 heart sounds can be considered non-stationary in comparison to heart sounds \cite{hadjileontiadis1998wavelet,hossain2003overview}. A modified version of wavelet thresholding is used to separate stationary sounds (lung sounds) from non-stationary sounds (heart sounds) using an adaptive threshold with adjusting multiplicative factor x \cite{hadjileontiadis1998wavelet,hossain2003overview,gradolewski2014wavelet}. Adjusting multiplicative factors of 2, 2.5, 2.7, 3, 3.5 and 4 were tested. 
    % \cite{hadjileontiadis1998wavelet,hossain2003overview,gradolewski2014wavelet, messer2001optimal}
    \\
    \hline
    Wavelet Decomposition and Singular Spectrum Analysis (n)
    & Wavelet decomposition is performed to obtain 0-500\,Hz signal. Singular spectrum analysis is then performed to obtain just heart sounds \cite{mondal2017noise}. 
    \\
    \hline 
    Non-Local Means
    & Identifies and averages similar heart and lung sound components \cite{rudnitskii2014using}. Due to the high computational cost of the method on our dataset, we did not implement this method in this study.
    %Non-local means identifies regions of the chest sound recording that are similar and assumed to be either heart or lung sound components. With the identified regions, they are then averaged, removing any potential noise and obtaining clean heart and lung sound components \cite{rudnitskii2014using}. Due to the high computational cost of the method on our dataset, we did not implement this method in this study. 
    \\
    \hline
    Autoencoder
    &  Encoder decomposes signal and temporal clustering is performed to identify heart and lung components. These components are placed into the decoder to obtain heart and lung sounds \cite{tsai2020blind}. This method was not implemented as the autoencoder needs to be retrained successfully and properly with a large and diverse set of neonatal chest sound recordings instead of adult chest sounds.
    %Tsai et al. \cite{tsai2020blind} trained a deep autoencoder, which contains an encoder and a decoder to be able to reproduce the input of an adult heart-lung sound mixture as output. Using the encoder, chest sound recording can be represented by several features, each with a particular temporal activation. Temporal clustering is then performed to separate into either heart or lung components \cite{tsai2020blind,lin2017improving}. These feature components are then inputted into the decoder to obtain heart and lung sounds. This method was not implemented as the autoencoder needs to be retrained successfully and properly with large and diverse set of neonatal chest sound recordings instead of adult chest sounds.
    \\
    \hline
    \multicolumn{2}{l}{*Letter reference is used in Section~\ref{sec:results}, Figure~\ref{fig:artificial_results} and Table~\ref{tab1:regressionresults} as shorthand for method names.}\\    
    \end{tabular}
    \label{tab1:existing_methods}
\end{center}    
\end{table*} 

\section{Introduction}
% Special Issue Topic:
% Innovations in Wearable, Implantable, Mobile, & Remote Healthcare with IoT & Sensor Informations

% Focus should be remote and mobile healthcare with digital stethoscope sensor. 
% Real-time processing 
% Patient monitoring 
% Enhancing healthcare and social well-being with IoT assistive wearable, implantable, mobile, and remote healthcare informatic
% Future of healthcare informatics with big data and artificial intelligence

% topic sentence and linking sentence at the end 

\IEEEPARstart{A}{ccurate} and timely assessment for signs of serious health problems such as cardio-respiratory diseases is an essential requirement to provide care to newborns \cite{Newborns19:online}. Recording chest sounds with a stethoscope is a common and simple method to obtain such information. In recent times, the availability of digital stethoscopes for neonates has attracted several studies \cite{ramanathan2019digital,zhou2020acoustic,ramanathan2020assessment}. Existing digital stethoscopes can connect with smartphones for mobile health and can enable remote healthcare through telehealth. However, a higher level of noise in neonatal intensive care in comparison to adult and paediatric wards has resulted in poor quality chest sound recordings and inaccurate assessment. For instance, estimation of heart rate and breathing rate from low-quality signals is error-prone \cite{grooby2020neonatal,grooby2021real}. 

%\IEEEPARstart{A}{ccurate} and timely assessment for signs of serious health problems such as cardio-respiratory diseases is an essential requirement to provide care to newborns \cite{Newborns19:online}. Recording chest sounds with a stethoscope is a common and simple method to obtain such information. In recent times, the availability of digital stethoscopes for neonates has attracted several studies \cite{ramanathan2019digital,kevat2017digital,zhou2020acoustic,ramanathan2020assessment}. Existing digital stethoscopes can connect with smartphones for mobile health and can enable remote healthcare through telehealth. However, a higher level of noise in neonatal intensive care in comparison to adult and paediatric wards has resulted in poor quality chest sound recordings and inaccurate assessment. For instance, estimation of heart rate and breathing rate from low-quality signals has been shown to be error-prone \cite{grooby2020neonatal,grooby2021real}. 

Noise interference in chest sounds can be broken up into four groups. Firstly, there are external noises, such as crying, stethoscope movement, talking and general background noise in neonatal intensive care \cite{lahav2015questionable}. Secondly, while heart and lung sounds are both diagnostically important, they act as noise sources for one another. Thirdly, other internal body sounds such as bowel sounds, gastric reflux, and air swallowing. Finally, neonates, particularly ones born earlier than 32 weeks, can experience respiratory-related conditions, requiring respiratory support such as high flow, ventilator or bubble Continuous Positive Airway Pressure (CPAP), which are major sources of interference. Overall, it is essential to reduce these noises, and separate heart and lung sounds before any assessment and diagnosis. 
%Noise interference in chest sounds can be broken up into four groups. Firstly, there are external noises, such as crying, stethoscope movement, talking and general background noise in neonatal intensive care \cite{lahav2015questionable,livera2008spectral}. Secondly, chest sounds are typically a mixture of both heart and lung sounds. Both sounds are diagnostically important but act as noise sources for one another. Thirdly, other internal body sounds such as bowel sounds, gastric reflux, and air swallowing may be present in chest sounds. Finally, neonates, particularly ones born earlier than 32 weeks, can experience respiratory-related conditions, requiring respiratory support such as high flow, ventilator or bubble Continuous Positive Airway Pressure (CPAP), which are major sources of interference. Overall, it is essential to reduce external, internal, and respiratory support noises, and separate heart and lung sounds before any assessment and diagnosis. 

Denoising and sound separation methods to obtain high-quality heart and lung sounds can be broken up into multi-channel and single-channel methods. In multi-channel methods, a reference signal such as an electrocardiogram or secondary microphone placed either to capture external noise and/or secondary chest recording is used \cite{nersisson2017heart}. Overall, these methods require additional sensors, which are not always accessible and feasible to implement.

\IEEEpubidadjcol

For single-source sound separation and denoising, current methods have proven only partially effective. Table~\ref{tab1:existing_methods} summarises existing single-source sound separation methods and Section~\ref{sec:results} compares them. Many of the methods utilise heart sound segmentation to obtain S1 and S2 peaks. As shown in our past works, heart sound segmentation accuracy drops significantly for low-quality recordings \cite{grooby2020neonatal,grooby2021real}. Overall this means many of these methods are only effective in low-noise scenarios. For lung sound separation, this has proven difficult due to its large frequency band that overlaps with heart and noise sources. 

Another limitation with past works is that they rely on adult-based parameters, which are not suitable for the separation of neonatal chest sounds. To address this, existing methods that rely on heart sound segmentation, utilise a modified version suitable for neonates, which was developed in our past work \cite{grooby2020neonatal}. Additionally, instead of a single-value parameter being utilised, a set of relevant parameters suitable for neonatal chest sound separation are considered and tested, as highlighted in Table~\ref{tab1:existing_methods}. Finally, for singular spectrum analysis (l) specifically, an additional constraint of top eigenvalue pairs having the strongest frequency component less than 250\,Hz are assigned as heart components is added. This constraint was added to address the issue of misclassification of lung components as heart, as they also contain semi-periodic components. 

%For single-source sound separation and denoising, current methods have proven only partially effective. A common approach is to first obtain a reference heart sound signal through methods such as thresholding, singular spectrum analysis, adaptive filtering or hidden semi-Markov model \cite{pourazad2003heart,pourazad2006heart,yadollahi2006robust,moussavi2004heart,flores2007heart,ghaderi2011localizing,liang1997heart, schmidt2010segmentation, springer2015logistic}, and then perform similar processes as in multi-channel methods \cite{ghaderi2011localizing, gnitecki2007separating, tsalaile2007separation}. A key limitation to this approach is the accuracy of obtaining reference heart sound is dependent on relatively low noise content and generally fixed-parameter based on heart sound properties of periodicity, time intervals or frequency information of adults. 

%Generally, heart sound has proven easier to obtain either due to its periodic or relatively narrow band low-frequency nature, which has been exploited in the methods. Whereas, lung sounds have a large frequency band that overlaps with heart and noise sources. Additionally, for neonates, lung sounds are not reliably periodic making them harder to detect. Section~\ref{sec:existing_methods} reviews in detail existing single-source sound separation methods and Section~\ref{sec:results} compares them. 

This paper presents a new approach, based on Non-negative Matrix Factorisation (NMF) and Non-negative Matrix Co-Factorisation (NMCF) to obtain high-quality heart and lung sounds. The model is trained with high-quality heart, lung and noise sound examples. In NMF this training occurs beforehand, whereas in NMCF this training occurs in parallel with separating sounds from the noisy recording into heart, lung and noise. The training set enables the utilisation of more detailed information about the frequency and temporal aspects of heart, lung and noise. Whereas for NMCF, the parallel training and separation of sounds from the noisy recordings enable the model to adapt more specifically to that recording. Overall, this method enables higher quality lung and heart sounds to be generated for analysis.

A preliminary version of the NMCF work has been reported, which utilised high-quality heart and lung sounds, with no reference noise sound examples \cite{grooby2021new}. Initial results on a real-world dataset showed it was superior to existing NMF methods \cite{grooby2021new}.

Four key contributions are presented in this paper. First, existing single-source denoising and heart and lung sound separation methods originally developed for adults were adapted and implemented on newborn chest sound recordings. Second, new NMF and NMCF approaches focused on obtaining high-quality heart and lung sounds specifically for the newborn population are proposed. Third, we incorporate a noise component in the NMF and NMCF models, to separate the sounds into not only heart and lung sounds but also noise sounds. Finally, the methods are assessed using artificial and real-world noisy neonatal chest sounds with heart and lung signal quality, and heart and breathing rate accuracy.

The rest of this paper is organised as follows. Section~\ref{sec:NMCF} provides the background of proposed NMF and NMCF methods. Section~\ref{sec:methods} presents a detailed implementation of the methods and their evaluation. Results and discussion are provided in Sections~\ref{sec:results} and \ref{sec:discussion}, respectively. Section~\ref{sec:conclusion} concludes the whole work.

\section{Non-Negative Matrix Factorisation and Non-Negative Matrix Co-Factorisation}
\label{sec:NMCF}
NMF decomposes a given non-negative matrix $V\in\Re^{F \times T}_+$ into two non-negative matrices $W\in\Re^{F \times K}_+$ and $H\in\Re^{K \times T}_+$ \eqref{eq:V}, where $K<min(F,T))$ and E represents the reconstruction error between V and WH. 

\begin{equation}
\label{eq:V}
\begin{split}
V & =WH + E\\ 
%& =\Lambda + E
\end{split}
\end{equation}

In denoising and sound separation, V represents the magnitude of the time-frequency representation of the recording mixture \cite{grooby2021new}. %This representation is typically the short-time Fourier transform, but gammatone filterbank and q-transform (log-frequency) representations have also been explored \cite{shah2014blind, canadas2017non, gao2012unsupervised, lin2013blind}. 

The basis matrix, W contains the basis column vectors $w_1$ to $w_K$ that represent the spectral pattern of different types of signals sources (e.g. heart, lung and noise) or their sub-components. The activation matrix H, contains the temporal activation row vectors $h_1$ to $h_k$, that represents when the signal sources occur during a particular time frame. These sub-components can be combined such that the first set of components (1 to $b_h$), second set of components ($b_h+1$ to $b_h+b_l$) and third set of components ($b_h+b_l+1$ to $b_h+b_l+b_n$) represent heart ($V_h=W_hH_h$), lung ($V_l=W_lH_l$) and noise respectively ($V_n=W_nH_n$) \cite{grooby2021new}. 

For supervised and semi-supervised NMF, the basis matrix W is optimised with reference heart, lung and noise sounds during the training phase. The basis matrix W is then fixed and H is optimised with the noisy mixture recording during the test phase. In NMCF, instead of having a training and test phase, the basis matrix W is optimised simultaneously with sound separation. This method enables more efficient sound separation as the mixture recording can also contribute to the training of W \cite{grooby2021new}.

%In NMCF, instead of having a training and test phase, which occurs in supervised (Eq.\ref{eq:nmcfs}) and semi-supervised (Eq.\ref{eq:nmcfsemi}) NMF, the matrix basis matrix W is optimised simultaneously with sound separation. This method enables more efficient sound separation as the mixture recording can also contribute to the training of W \cite{de2020wheezing}.

%\begin{equation}
%\label{eq:nmcfs}
%\begin{split}
%W_h & =\min\limits_{W_h,H_h}(D(V_h|\hat{W_h}H_h))\\
%W_l & =\min\limits_{W_l,H_l}(D(V_l|\hat{W_l}H_l))\\
%H_m & =\min\limits_{H_m}(D(V_m|\hat{W}H_m))\\
%Where: & \quad H_m = [H_{mh};H_{ml}], \hat{W}=[\hat{W}_{h},\hat{W}_{l}]\\
%\end{split}
%\end{equation}

%\begin{equation}
%\label{eq:nmcfsemi}
%\begin{split}
%W_h & =\min\limits_{W_h,H_h}(D(V_h|\hat{W_h}H_h))\\
%H_m & =\min\limits_{W_l,H_m}(D(V_m|\hat{W}H_m))\\
%Where: & \quad H_m = [H_{mh};H_{ml}], \hat{W}=[\hat{W}_{h},\hat{W}_{l}]\\
%\end{split}
%\end{equation}

\begin{figure*}[t]
\centering
\includegraphics[scale=0.13,trim={0.5cm 0.5cm 0.5cm 0.5cm}]{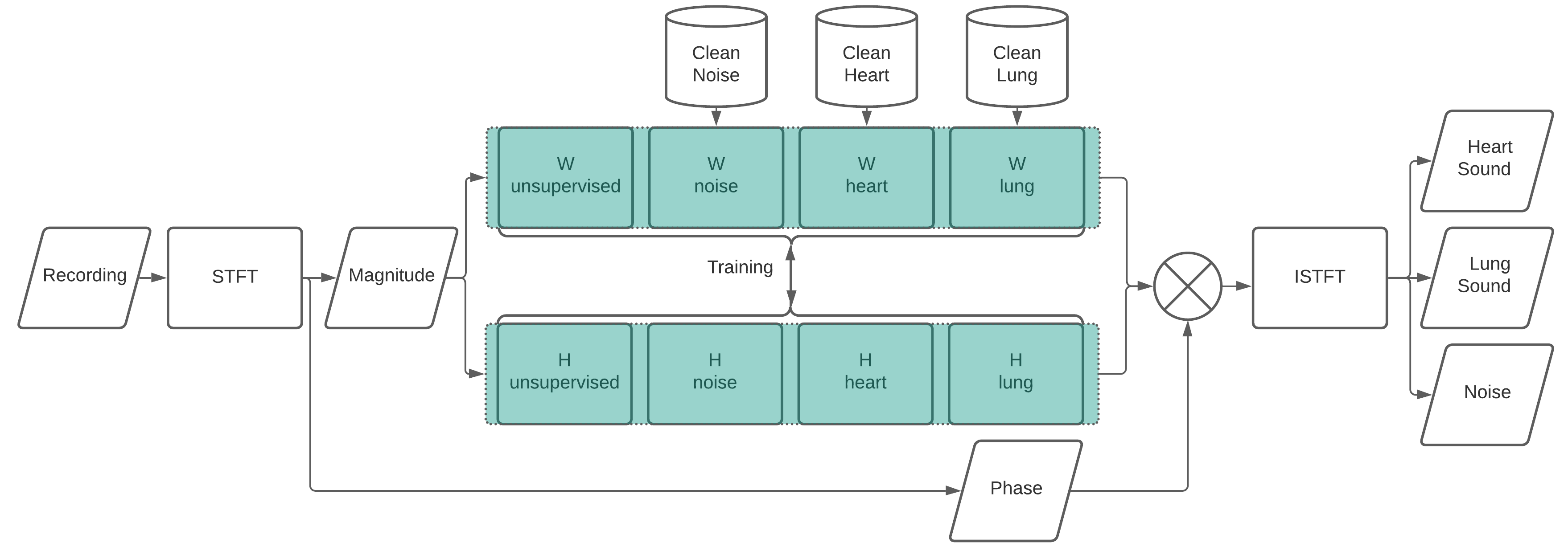}
\caption{Proposed NMCF Flowchart}
\label{fig:proposed_nmcf_flowchart}
\end{figure*}

However, as obtaining pure heart or lung sounds is not feasible, we propose a modified version of NMF and NMCF (equations \eqref{eq:mnmf} and \eqref{eq:mnmcf}, respectively) as described in  Algorithm~\ref{alg:nmcf}, and Figure~\ref{fig:proposed_nmcf_flowchart}. In this version, datasets of high-quality heart, lung, and noise sounds  are used in the cost function to enable the generalisation of $W_h$, $W_l$ and $W_n$ respectively for the sound separation. Note, these reference datasets are generalisable and are not obtained from the same subject as the noisy mixture recording that is being denoised. 

In both the proposed NMF and NMCF equations \eqref{eq:mnmf} and \eqref{eq:mnmcf}, the matrices W and H are optimised by minimising the cost function D \eqref{eq:sparsity}. $D_\beta$ refers to $\beta$-divergence cost function, with the most popular values being $\beta =0,1,2$ \cite{grooby2021new}. A sparsity penalty on the activation matrix H is calculated based on the L1-norm of H, and $\mu$ controls the importance of the sparsity constraint. The sparsity penalty enables more detailed decomposition both temporally and spectrally while ensuring only a small set of meaningful basis vectors are active at a single time frame \cite{le2015sparse}. 

%In both the proposed NMF and NMCF equations (Eq.~\ref{eq:mnmf}) and \ref{eq:mnmcf}), the matrices W and H are optimised by minimising the cost function D (Eq.~\ref{eq:sparsity}). The most popular cost functions for $D_\beta$ as shown in Eq.~\ref{eq:d} are; $\beta =0$ as Ikaura-Saito distance, $\beta =1$ as Kullback-Leibler divergence and $\beta =2$ which yields Euclidean distance \cite{grooby2021new}. A sparsity penalty on the activation matrix H is calculated based on the L1-norm of H, and $\mu$ controls the importance of the sparsity constraint. The sparsity penalty enables more detailed decomposition both temporally and spectrally while ensuring only a small set of meaningful basis vectors are active at a single time frame \cite{le2015sparse}. 
For proposed NMF \eqref{eq:mnmf} we have introduced a cost function that utilises datasets of clean heart, lung and noise sounds that are obtained from different subjects than the noisy mixture recording. This differs from past work which either requires simultaneous reference recordings from the same subject or relies on blind decomposition \cite{nersisson2017heart,grooby2021new}. 

\begin{equation}%[tb]
\label{eq:mnmf}
\begin{split}
W_h & =\min\limits_{W_h,H_h}(\sum_{ih=1}^{eh} D(V_h^{(ih)}|\hat{W_h}H_h^{(ih)}))\\
W_l & =\min\limits_{W_l,H_l}(\sum_{il=1}^{el} D(V_l^{(il)}|\hat{W_l}H_l^{(il)}))\\
W_n & =\min\limits_{W_n,H_n}(\sum_{in=1}^{en} D(V_n^{(in)}|\hat{W_n}H_n^{(in)}))\\
H_m & =\min\limits_{W_{un},H_m}(D(V_m|\hat{W}H_m))\\
Where: & \quad H_m = [H_{mh}; H_{ml};  H_{mn}; H_{mun}],\\ 
& \hat{W}=[\hat{W}_{h}, \hat{W}_{l}, \hat{W}_{n}, \hat{W}_{un}]\\
\end{split}
\end{equation}

While for proposed NMCF \eqref{eq:mnmcf}, building on our past work, we have introduced a supervised noise component \cite{grooby2021new}. The weighting factors $\lambda_h$, $\lambda_l$ and $\lambda_n$ represent the level of co-factorisation and are treated as hyperparameters. Additionally, an unsupervised component $W_{un}$ is added to deal with the large variety of noises that are not covered, therefore avoiding these components being assigned to the heart or lung components. 
\begin{equation}
\label{eq:mnmcf}
\begin{split}
W,H_m & =\min\limits_{W,H_m,H_h,H_l,H_n}(\lambda_h D(V_m|\hat{W_h}H_{mh})\\
&+ \lambda_l D(V_m|\hat{W_l}H_{ml})
+\lambda_n D(V_m|\hat{W_n}H_{mn})\\
&+ D(V_m|\hat{W_{un}}H_{mun})\\
&+\frac{1}{eh}\sum_{ih=1}^{eh}  D(V_h^{(ih)}|\hat{W_h}H_h^{(ih)})\\
&+\frac{1}{el}\sum_{il=1}^{el} D(V_l^{(il)}|\hat{W_l}H_l^{(il)})\\
&+\frac{1}{en}\sum_{in=1}^{en} D(V_n^{(in)}|\hat{W_n}H_n^{(in)}))\\
Where: & \quad H_m = [H_{mh}; H_{ml};  H_{mn}; H_{mun}],\\ 
& \hat{W}=[\hat{W}_{h}, \hat{W}_{l}, \hat{W}_{n}, \hat{W}_{un}]\\
\end{split}
\end{equation}

%\begin{equation}
%\label{eq:mnmcf}
%\begin{split}
%W,H_m & =\min\limits_{W,H_m,H_h,H_l,H_n}(D(V_m|\hat{W}H_m)\\
%&+\sum_{ih=1}^{eh} \lambda^{(ih)}_h D(V_h^{(ih)}|\hat{W_h}H_h^{(ih)})\\
%&+\sum_{il=1}^{el} \lambda^{(il)}_l D(V_l^{(il)}|\hat{W_l}H_l^{(il)})\\
%&+\sum_{in=1}^{en} \lambda^{(in)}_n D(V_n^{(in)}|\hat{W_n}H_n^{(in)}))\\
%Where: & \quad H_m = [H_{mh}; H_{ml};  H_{mm}; H_{mun}],\\ 
%& \hat{W}=[\hat{W}_{h}, \hat{W}_{l}, \hat{W}_{n}, \hat{W}_{un}]\\
%\end{split}
%\end{equation}

\begin{equation}
\label{eq:sparsity}
\begin{split}
Where: &\quad D(V|\hat{W}H) =D_\beta(V|\hat{W}H) + \mu||H||_1,\\
Where: &\quad \hat{W}= [\frac{w_1}{||w_1||},\frac{w_2}{||w_2||},...,\frac{w_K}{||w_K||}]
\end{split}
\end{equation}

\begin{comment}
\begin{equation}
\label{eq:d}
D_\beta(x|y)=\begin{cases}
			\frac{(x^\beta-y^\beta-\beta y^{\beta-1}(x-y))}{\beta(\beta-1)}, & \text{if $\beta \in \Re \backslash \{0,1\}$}\\
            x(log(x)-log(y))+(y-x), & \text{if $\beta=1$}\\
            \frac{x}{y}-log(\frac{x}{y})-1, & \text{if $\beta=0$}\\
		 \end{cases}
\end{equation}
\end{comment}

Based on the cost function in \eqref{eq:sparsity}, the multiplicative update rule for W and H are shown in \eqref{eq:wupdate} and \eqref{eq:hupdate} respectively. Note that division and $\otimes$ refer to element-wise division and multiplication. 

\begin{equation}
\label{eq:wupdate}
\begin{split}
W & \leftarrow \hat{W} \otimes\\ &\frac{(\Lambda^{\beta-2}\otimes V)H^T+\hat{W} \otimes (11^T(\hat{W} \otimes (\Lambda^{\beta-1}H^T)))}
{\Lambda^{\beta-1}H^T+\hat{W} \otimes (11^T(\hat{W} \otimes ((\Lambda^{\beta-2}\otimes V)H^T)))}\\
Where: &\quad \Lambda = WH, \quad 1 =\text{length F column vector of ones}\\
W & \leftarrow W \otimes \frac{W_{num}(V,W,H)}{W_{dem}(V,W,H)}
\end{split}
\end{equation}

\begin{equation}
\label{eq:hupdate}
\begin{split}
H &\leftarrow H \otimes \frac{\hat{W}^T(V\otimes\Lambda^{\beta-2})}{\hat{W}^T\Lambda^{\beta-1}+\mu}, \Lambda = WH\\
H & \leftarrow H \otimes \frac{H_{num}(V,W,H)}{H_{dem}(V,W,H)}
\end{split}
\end{equation}

\begin{algorithm}[ht]
\begin{flushleft}
\caption{Proposed NMCF}
\label{alg:nmcf}
\begin{algorithmic}[1]
	\STATE $V_m,Phase=stft(audio_m)$\;
	\STATE $V_h^{(ih)}=stft(audio_h^{(ih)})$\;
	\STATE $V_l^{(il)}=stft(audio_l^{(il)})$\;
	\STATE $V_n^{(in)}=stft(audio_n^{(in)})$\;
	\STATE $init \quad H_{mh},H_{ml},H_{mn},H_{mun},H_h^{ih}, H_l^{il}, H_n^{in}$\;
	\STATE $init \quad \hat{W}_{h}, \hat{W}_{l}, \hat{W}_{n}, \hat{W}_{un}$\;
    \STATE $W=[\hat{W}_{h}, \hat{W}_{l}, \hat{W}_{n}, \hat{W}_{un}]$\;
	\STATE $H_{m}=[H_{mh};H_{ml};H_{mn};H_{mun}]$\;
	\FOR{i=1:maxiter}
	
	\STATE $H_m \leftarrow H_m \otimes \frac{H_{num}(V_m,W,H_m)}{H_{dem}(V_m,W,H_m)}$
    
    \STATE $H_h^{ih} \leftarrow H_h^{ih} \otimes \frac{H_{num}(V_h^{ih},W_h,H_h^{ih})}{H_{dem}(V_h^{ih},W_h,H_h^{ih})}$\;
    
    \STATE $H_l^{il} \leftarrow H_l^{il} \otimes \frac{H_{num}(V_l^{il},W_l,H_l^{il})}{H_{dem}(V_l^{il},W_l,H_l^{il})}$\;
    
    \STATE $H_n^{in} \leftarrow H_n^{in} \otimes \frac{H_{num}(V_n^{in},W_n,H_n^{in})}{H_{dem}(V_n^{in},W_n,H_n^{in})}$\;
	
    \STATE $W_{h} \leftarrow W_h \otimes 
    \frac{\lambda_h W_{num}(V_m,W_{h},H_{mh})
    +\frac{1}{eh}\sum_{ih=1}^{eh} W_{num}(V^{(ih)},W_{h},H_h^{(ih)})}
    {\lambda_h W_{dem}(V_m,W_{h},H_{mh})
    +\frac{1}{eh}\sum_{ih=1}^{eh} W_{num}(V^{(ih)},W_{h},H_h^{(ih)})}$\;
    
    \STATE $W_h = normalisation(W_h)$\;
            
    \STATE $W_{l} \leftarrow W_l \otimes 
    \frac{\lambda_l W_{num}(V_m,W_{l},H_{ml})
    +\frac{1}{el}\sum_{il=1}^{el} W_{num}(V^{(il)},W_{l},H_l^{(il)})}
    {\lambda_l W_{dem}(V_m,W_{l},H_{ml})
    +\frac{1}{el}\sum_{il=1}^{el} W_{num}(V^{(il)},W_{l},H_l^{(il)})}$\;
    
    \STATE $W_l = normalisation(W_l)$\;
    
    \STATE $W_{n} \leftarrow W_n \otimes 
    \frac{\lambda_n W_{num}(V_m,W_{n},H_{mn})
    +\frac{1}{en}\sum_{in=1}^{en} W_{num}(V^{(in)},W_{n},H_n^{(in)})}
    {\lambda_n W_{dem}(V_m,W_{n},H_{mn})
    +\frac{1}{en}\sum_{in=1}^{en} W_{num}(V^{(in)},W_{n},H_n^{(in)})}$\;
    
    \STATE $W_n = normalisation(W_n)$\;
    
    \STATE $W_{un} \leftarrow W_{un} \otimes 
    \frac{W_{num}(V_m,W_{un},H_{mun})}
    {W_{dem}(V_m,W_{un},H_{mun})}$\;
    
    \STATE $W_{un} = normalisation(W_{un})$\;
	
\ENDFOR
    \STATE $mask_h =\frac{\hat{W}_{h}H_{mh}}{WH}$\;
	\STATE $mask_l =\frac{\hat{W}_{l}H_{ml}}{WH}$\;
	\STATE $V_h = np.multiply(V_m, mask_h)$\;
	\STATE $V_l = np.multiply(V_m, mask_l)$\;
	\STATE $audio_{heart}=istft(V_h,Phase)$\;
	\STATE $audio_{lung}=istft(V_l,Phase)$\;
\end{algorithmic}
\end{flushleft}
\end{algorithm}

\section{Methods}
\label{sec:methods}

\subsection{Data Acquisition and Preprocessing}

% Without respiratory support
The study was conducted at Monash Newborn, Monash Children’s Hospital. It was approved by the Monash Health Human Research Ethics Committee (HREA/18/MonH/471). Recordings were obtained from the right anterior chest of preterm and term newborns using a digital stethoscope \cite{zhou2020acoustic,ramanathan2020assessment, grooby2021real}. A subset of these recordings had synchronous vital signs from electrocardiogram for reference heart and breathing rate which was used in Section~\ref{sec:HRBR error}. The breakdown of the recordings is shown in Table~\ref{tab1:data}.

\begin{table}[ht]
    \begin{center}
\caption{Data Breakdown}
    \begin{tabular}{ |m{4cm}||M{3cm}|}
    \hline
    \textbf{Type}
    & \textbf{Details}
    \\
    \hline
    Without Respiratory Support
    & 318 60\,s recordings
    \\
    \hline
    With Respiratory Support
    & 79 60\,s recordings
    \\
    \hline
    Without Respiratory Support \newline with Synchronous Vital Signs
    & 22 60\,s recordings
    \\
    \hline
    With Respiratory Support \newline with Synchronous Vital Signs
    & 9 60\,s recordings
    \\
    \hline
    \end{tabular}
    \label{tab1:data}
\end{center}    
\end{table} 

\subsection{Reference Sounds}
\label{sec:reference_sounds}

\begin{figure}[tp]
\centering
\includegraphics[scale=0.28,trim={0.5cm 3cm 0.5cm 0.5cm}]{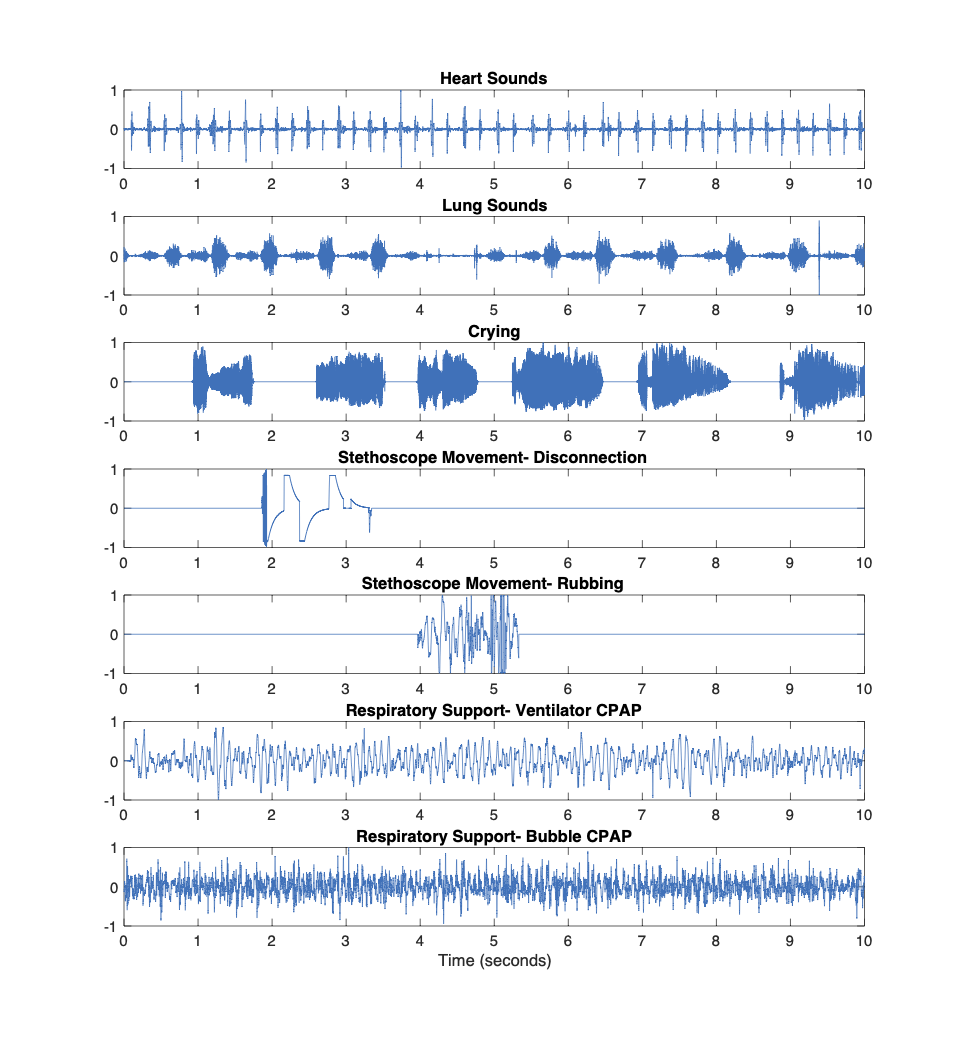}
\caption{Example Reference Sounds}
\label{fig:reference_sounds}
\end{figure}

Reference heart, lung, crying, stethoscope movement, and respiratory support sounds were required for two purposes. Firstly, several existing methods and the proposed NMF and NMCF methods required reference high-quality sounds to enable sound separation either through training or comparison during the sound separation procedure itself. Secondly, these reference sounds were used to create artificial mixtures to enable evaluation of the sound separation methods as described in Section~\ref{sec:BSS_eval}. Example reference sounds are shown in Figure~\ref{fig:reference_sounds}. 

Heart and lung sounds were obtained from the recordings of newborns without respiratory support. As pure heart and lung sounds are required to construct an artificial dataset, these recordings were 4\textsuperscript{th}-order Butterworth bandpass filtered with passband frequencies 50-250\,Hz and 200-1000\,Hz, to separate heart and lung sounds, respectively \cite{grooby2020neonatal,nersisson2017heart}.

The filtered recordings were then annotated by 3 clinicians and 4 electrical engineers familiar with biomedical auscultation for heart and lung signal quality on a 5-level scale. The score of 1 referred to only noisy and hardly detectable heartbeats/breathing periods, and 5 referred to clear heart/lung sounds with little to no noise. Mean annotated scores of 4 and above were then assessed visually and through audio, and only the recordings with strong heart/lung sounds and little to no noise were chosen as reference signals. In total, 17 signals (9 subjects) with 7 (3 subjects) having synchronous heart rate remained for reference heart sounds and 9 signals (7 subjects) with no synchronous breathing rate remained for reference lung sounds. 

To obtain the cry sounds, the cry detection algorithm implemented in previous work was used \cite{grooby2020neonatal}. Regions containing at least 10\,s of crying were then determined and extracted. These segments were then
2\textsuperscript{nd}-order Butterworth high-pass filtered with cutoff 300\,Hz to remove heart sounds. Finally, regions not containing crying such as inhale and other lung sounds were replaced with zeros. In total 42 signals (41 subjects) were obtained. 

Two clinicians and 1 electrical engineer familiar with biomedical auscultation manually annotated recordings for presence and volume (low, medium or high) of stethoscope movement and respiratory support noise. Stethoscope movement noise can be typically characterised as a short <2\,s burst of noise. These sections of stethoscope movement were isolated and classified as either disconnection or rubbing noise. Disconnection noise only has the stethoscope movement sound present and not heart nor lung sounds, as the stethoscope has disconnected from the newborn's chest. Whereas for rubbing noise, the stethoscope is still in connection with the newborn's chest and can potentially still contain heart and lung sounds. In total 18 signals (12 subjects) of varying lengths were obtained. To ensure all signals were of length 10\,s, zeros were appended to the beginning and end of the signals. The ratio of the total number of zeros placed at the beginning or end of the signals was randomly determined, to ensure that stethoscope movement occurs at a random position in the 10\,s signals. 

Recordings annotated with a high level of respiratory support noise and next to no presence of other sounds were chosen. From this, 2 signals (1 subject) of bubble CPAP and 5 signals (1 subject) of ventilator CPAP were obtained. Given the difficultly of obtaining pure support sounds from chest sound recordings, pure respiratory support sounds were also collected by placing the digital stethoscope on the respiratory support machine or tubing. An additional 3 signals for both bubble and ventilator CPAP were obtained from the respiratory support machine itself. 

\subsection{Implementation}

For both proposed NMCF and NMF, the number of bases used is 20 for heart, lung and all noise sounds each. For real-world sounds, an unsupervised noise component is also added with the number of bases equaling 10. At max 10 examples of heart, lung and noise sounds were used and 100 iterations of multiplicative update of activation and basis matrix. 

The following parameters for the proposed NMCF and NMF were tested:
\begin{itemize}
  \item Time-Frequency Representation: 
  \begin{itemize}
    \item Gammatone filterbank \cite{gao2012unsupervised} with a frequency range of 50-2000\,Hz and frequency bin size of either 128, 256, 512 or 1028. 
    \item Q-transform (log-frequency) with 64 frequency bins per octave
    \item Short-Time Fourier Transform (STFT) with Fast Fourier Transform (FFT) Size = 2048 samples, Window Size = 512 samples and Hop Size = 256 samples similar to previous work \cite{grooby2021new}, or, FFT Size = 1024 samples, Window Size = 512 samples and Hop Size = 256 samples. 
  \end{itemize}
  \item Sparsity penalty ($\mu$): 0, 0.01 and 0.1 
  \item Beta loss ($\beta$): 0, 1 and 2
  \item NMCF $\lambda_h$, $\lambda_l$ and $\lambda_n$: 0, 0.25, 0.5, 0.75 and 1
\end{itemize}

For the artificial dataset, subject-wise cross-validation was used to generate 7 folds of artificial mixtures. One fold was used for hyperparameter optimisation for the proposed NMCF and NMF methods and existing sound separation methods. The remaining six-folds were used for evaluation of the sound separation, with results shown in Figures~\ref{fig:artificial_heart} and \ref{fig:artificial_lung}. 

Based on the hyperparameter optimisation results, STFT with FFT size=1024 samples, window size=512 samples and hop size=256 samples, sparsity penalty=0.1 and beta loss=1 were used for both proposed NMCF and NMF methods to separate the real-word database recordings. Additionally, $\lambda_h$=0, $\lambda_l$=0 and $\lambda_n$=0.25 (Stethoscope Movement Noise, Ventilator CPAP Noise) and $\lambda_n$=0.75 (Cry Noise, Bubble CPAP Noise) were used for the proposed NMCF method. 

\subsection{Performance Evaluation}
\subsubsection{Artificial Mixtures Evaluation}
\label{sec:BSS_eval}

Using the reference sounds obtained in Section~\ref{sec:reference_sounds}, artificial mixtures are generated. 
The most common method in literature for generating chest sound mixtures is the simple addition of all reference sounds, referred to as an instantaneous mixture \eqref{eq:instantaneous_mixture}. However, as chest sounds are not simple instantaneous mixtures of heart, lung, and noise sounds, a convolutive mixture model was also adopted \eqref{eq:convolutive_mixture} \cite{ghaderi2011localizing}. To achieve convolutive mixing, three randomly generated finite-impulse response (FIR) filters of length 4 were generated to mix heart ($a_{heart}$), lung ($a_{lung}$)and noise ($a_{noise}$). FIR filter coefficients $a_{heart}$, $a_{lung}$, $a_{noise}$ are scaled to have magnitude of 1 . 

\begin{equation}
\label{eq:instantaneous_mixture}
s_{mixture}(t) = s_{heart}(t) + s_{lung}(t) +s_{noise}(t)
\\
\end{equation}

\begin{equation}
\label{eq:convolutive_mixture}
\begin{split}
s_{mixture}(t)  = \sum_{k=0}^{3}(& a_{heart}(k)s_{heart}(t-k) +\\
& a_{lung}(k)s_{lung}(t-k) +\\
& a_{noise}(k)s_{noise}(t-k))
\\
\end{split}
\end{equation}

Before mixing heart, lung and noise sounds, they were scaled to achieve the desired signal-to-noise ratio. Heart sounds were first scaled to achieve a heart-to-lung sounds ratio of -10, -5, 0, 5, 10, 15 and 20\,dB \eqref{eq:snr_heart_lung}. Once scaled, heart and lung sounds were combined and then scaled to achieve chest-to-noise sounds ratio of -10, -5, 5, 0 and 10\,dB \eqref{eq:snr_chest_noise}. 

\begin{equation}
\label{eq:snr_heart_lung}
s_{chest}(t)=10^{factor/20}*s_{heart}(t)+s_{lung}(t)
\\
\end{equation}

\begin{equation}
\label{eq:snr_chest_noise}
s_{mixture}(t)=10^{factor/20}*s_{chest}(t)+s_{noise}(t)
\\
\end{equation}

Using the generated artificial mixtures and reference sounds used to create these mixtures, several signal quality metrics can be calculated using the blind source separation evaluation toolbox \cite{vincent2006performance,fevotte2005bss_eval}. With this toolbox, estimated heart, lung and noise sounds are decomposed into 4 components, namely; true reference sound ($s_{target}$), interference noise ($e_{inter}$), additive noise ($e_{noise}$), algorithmic artifact noise ($e_{artif}$). Once decomposed, signal-to-distortion ratio (SDR \eqref{eq:sdr}), signal-to-interference ratio (SIR \eqref{eq:sir}) and scale-invariant signal-to-distortion ratio (SI-SDR \eqref{eq:sdr}) are calculated. Both SDR and SI-SDR are overall metrics of signal quality. The difference between SDR and SI-SDR is that SDR uses a full 512-tap FIR filter whereas SI-SDR uses a single coefficient to account for allowable scaling discrepancies between estimated separated sounds and reference sounds \cite{vincent2006performance,le2019sdr}. Therefore SI-SDR harshly penalises temporal distortions and is only suitable for the evaluation of instantaneous mixtures in comparison to SDR.  

\begin{equation}
\label{eq:sdr}
SDR = 10log_{10}\frac{||s_{target}(t)||^2}
{||e_{inter}(t)+e_{noise}(t)+e_{artif}(t)||^2}
\\
\end{equation}

\begin{equation}
\label{eq:sir}
SIR = 10log_{10}\frac{||s_{target}(t)||^2}
{||e_{interf}(t)||^2}
\\
\end{equation}

\subsubsection{Heart Rate and Breathing Rate Error}
\label{sec:HRBR error}

A goal of obtaining high-quality heart and lung sounds is to achieve accurate heart and breathing rate estimates. These vital sign estimates are essential in cardio-respiratory health assessment, to enable proper clinical care to be determined and provided \cite{king2020tools,jain2019neonatal}. %For clinical use, both heart and breathing estimates should have a mean absolute error of less than 5 beats per minute \cite{springer2014robust,nizami2018measuring}. 

For the heart audio recordings, heart rate in beats per minute was estimated for each second with a sliding window of 3\,s. Heart rate was calculated using the modified version of the method by Springer et al.  \cite{springer2015logistic} as proposed in our past work \cite{grooby2021real}.

For lung audio recordings, breathing rate in breaths per minute was estimated every second with a sliding window of 6\,s. For breathing rate estimation, power spectral envelope is calculated for the frequency range 300-450\,Hz and then peak detection is performed \cite{grooby2020neonatal}. 

%For the heart audio recordings, heart rate in beats per minute was estimated every second with a sliding window of 3\,s. A sliding window of 3\,s was chosen as this is a sufficient length to obtain a minimum of 3 heartbeats, necessary for accurate heart rate estimation. Two methods were used to estimate heart rate. Firstly, using the method proposed by Schmidt et al. \cite{schmidt2010segmentation}, where the autocorrelation of the Hilbert Envelope is calculated. The maximum peak is then detected in the autocorrelation signal between the bounds of 70-220 beats per minute. The range of 70-220 beats per minute is chosen as this is the typical heart rate range for newborns \cite{grooby2020neonatal,FastSlow12:online,Normalhe33:online}. 
%The second method proposed by Springer et al. \cite{springer2015logistic} uses the initial estimate of heart rate from the Schmidt et al. method as input into a duration-dependent hidden Markov model, to segment the heartbeats into 4 states, namely S1, S2, systolic and diastolic.

%For lung audio recordings, breathing rate in breaths per minute was estimated every second with a sliding window of 6\,s. Similarly as before, a sliding window of 6\,s was chosen as this is a sufficient length to obtain a minimum of 3 breathing periods, which is necessary for accurate breathing rate estimation. For breathing rate estimation, power spectral envelope is calculated for the frequency range 300-450Hz and then peak detection is performed \cite{grooby2020neonatal}. 

\subsubsection{Signal Quality Assessment}
\label{sec:SQI}

An automated signal quality assessment method to classify real-world heart and lung signal quality on a 5-level scale was developed in our previous works \cite{grooby2020neonatal,grooby2021real}. A score of 1 referred to only noisy and hardly detectable heartbeats/breathing periods, and 5 referred to clear heart/lung sounds with little to no noise.  

To calculate the signal quality, up to 15 features for heart signal quality, and up to 20 features for lung signal quality, were extracted from chest sound recordings and used as input into the regression classifier. Features extracted included statistical features (variance, skewness, and kurtosis), predictive fitting coefficients, heart and lung segmentation quality and agreement, Mel-frequency coefficients, wavelet, entropy and power. 

The dataset used for training was a subset of the no respiratory support recordings. In total the regression classifier was trained on 206 recordings from 97 subjects for lung sound quality estimation and 223 recordings from 92 subjects for heart sound quality estimation. Reference quality annotations were provided by 3 clinicians and 4 electrical engineers familiar with biomedical auscultation\cite{grooby2020neonatal,grooby2021real}. Note that these recordings are from different subjects than the reference sounds used for training the sound separation methods and creation of the artificial dataset in Section~\ref{sec:reference_sounds}. 

\subsubsection{Real-Time Analysis}
\label{sec:real_time}

The median time for chest sound separation for an example 10\,s was calculated using MATLAB 2021a with MacBook Pro CPU 2.3\,GHz 8-Core Intel~i9. 

Computation cost per 10\,s recording is shown in Table~\ref{tab1:regressionresults}. The proposed NMCF method takes a median  of 28.2\,s and 28.3\,s with and without supervised decomposition components respectively. These computational times make the proposed NMCF method not suitable for real-time processing. For the proposed NMF method, the median computational time of 275\,ms and 342\,ms with and without supervised decomposition components are observed. As the computational times are less than 400\,ms, the proposed NMF method is suitable for real-time processing using the stated laptop specifications \cite{grooby2021new}. For existing methods, adaptive Fourier decomposition (c), adaptive line enhancement (d), filtering (f), interpolation (g) and modulation filtering (h) are suitable for real-time processing using the stated laptop specifications. 

\subsubsection{Statistical Analysis}

Statistical tests were performed to determine if the proposed NMF and NMCF methods are significantly outperforming existing methods. Using the Jarque-Bera test, artificial dataset signal quality improvement results were not normally distributed. Therefore, median values are reported and a one-sided Wilcoxon signed-ranked test was used to test significance in Section~\ref{sec:results}. Similarly, vital sign estimation error results were not normally distributed and thus one-sided Wilcoxon signed-ranked test was used to test significance. However, as median results were predominately zero, mean and standard deviation results were shown in Table~\ref{tab1:regressionresults} to be more informative. Whereas, signal quality improvement values for the real-world dataset were normally disturbed. Therefore, mean and standard deviation values are reported in Table~\ref{tab1:regressionresults} and a one-sided t-test was used to test significance in Section~\ref{sec:results}. 

\section{Results}
\label{sec:results}

\begin{figure*}
     \centering
     \begin{subfigure}[b]{1\textwidth}
         \centering
         \includegraphics[scale=0.33,trim={5.5cm 3cm 0.5cm 0.5cm}]{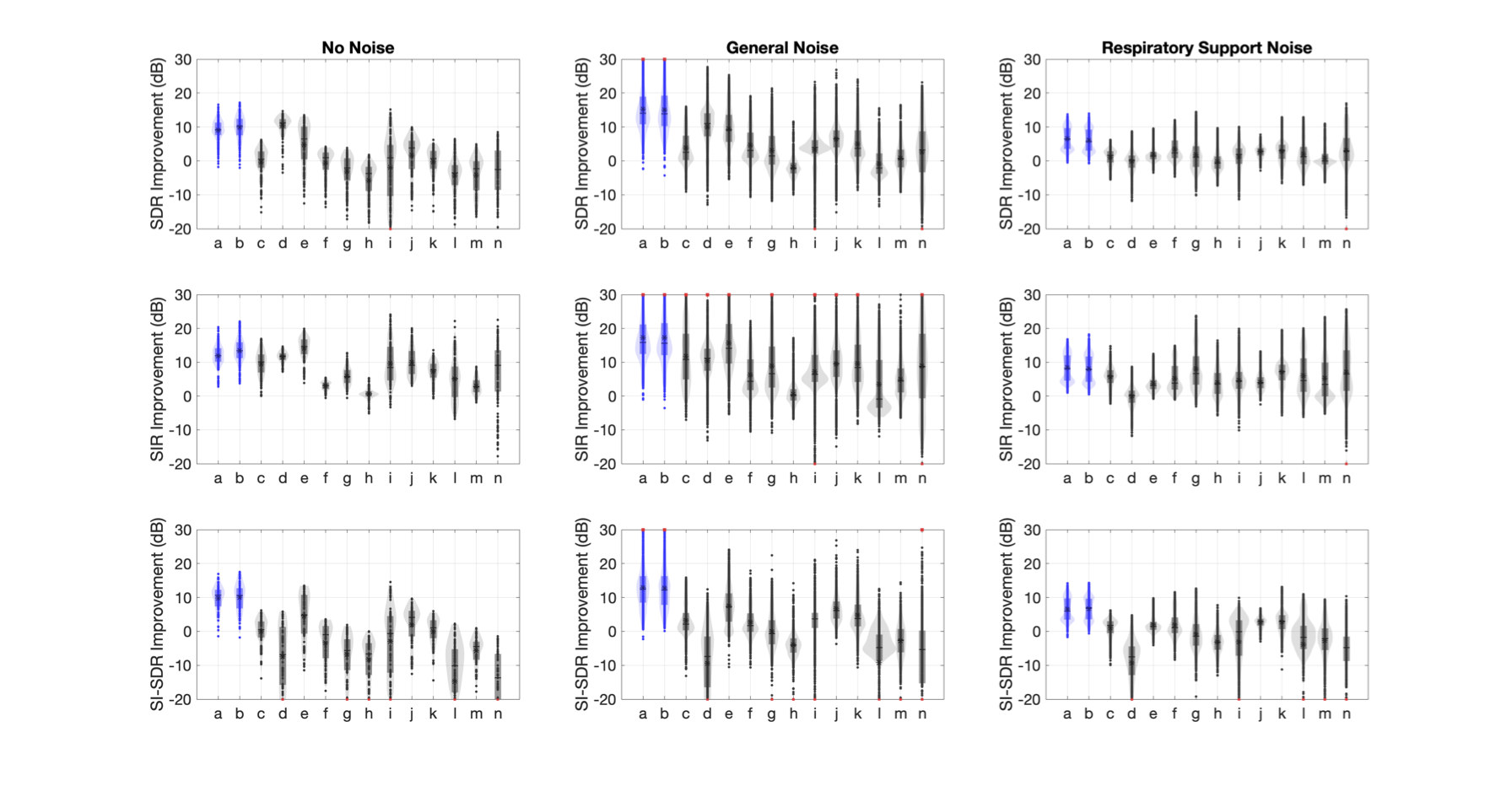}
         \caption{Heart}
         \label{fig:artificial_heart}
     \end{subfigure}
     \hfill
     \begin{subfigure}[b]{1\textwidth}
         \centering
         \includegraphics[scale=0.33,trim={5.5cm 3cm 0.5cm 0cm}]{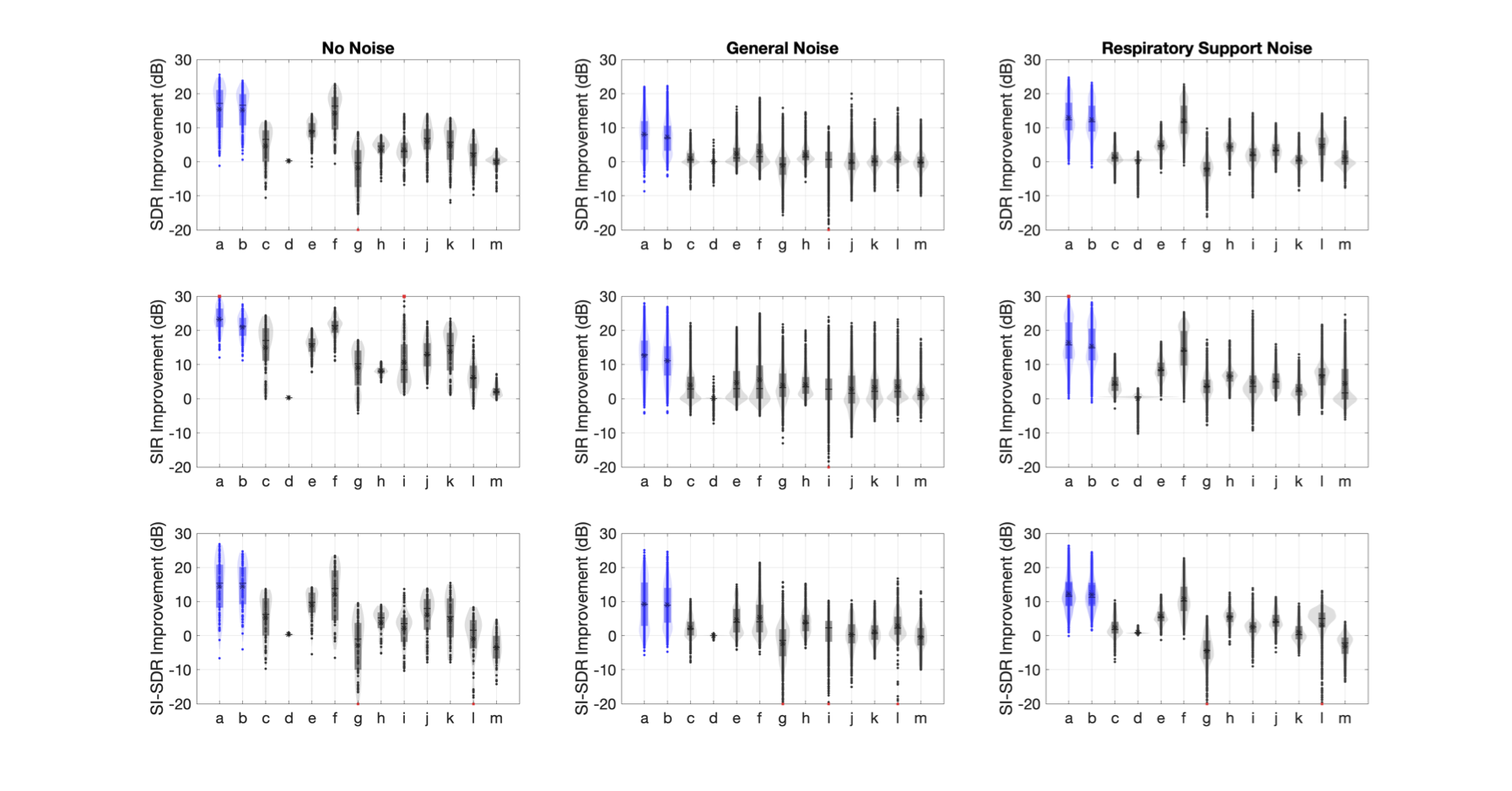}
         \caption{Lung}
         \label{fig:artificial_lung}
     \end{subfigure}
     \hfill
    \caption{Artificial Dataset- Sound Separation Results. The top 9 subplots are for heart sound separation (Figure~\ref{fig:artificial_heart}) and bottom 9 subplots are for lung sound separation (Figure~\ref{fig:artificial_lung}). SDR, SIR and SI-SDR improvement results are calculated according to Section~\ref{sec:BSS_eval}, and displayed using box and violin plots as described in Figure~\ref{fig:violin_plot}. No noise refers to only heart-lung sound mixtures, general noise refers to heart-lung sound and either cry or stethoscope movement noise mixtures, and respiratory support noise refers to heart-lung sound and either bubble or ventilator CPAP noise mixtures. Methods a and b (in blue) are proposed NMCF and NMF methods, and methods c to n (in black) are existing methods, as specified in Table~\ref{tab1:existing_methods}.}
    \label{fig:artificial_results}
\end{figure*}

Figure~\ref{fig:artificial_results} and Table~\ref{tab1:regressionresults} show the artificial and real-world chest sound separation results respectively. Methods (a), (a un), (b), (b un)  are proposed NMCF, NMCF with Unsupervised Components, NMF and NMF with Unsupervised components methods respectively. Methods (c) to (n) are existing methods, as specified in Table~\ref{tab1:existing_methods}.

%Figure~\ref{fig:artificial_results} and Table~\ref{tab1:regressionresults} show the artificial and real-world chest sound separation results respectively. Letters a to n refer to: (a) Proposed NMCF, (b) Proposed NMF, (c) Adaptive Fourier Decomposition \cite{wang2015adaptive}, (d) Adaptive Line Enhancement \cite{tsalaile2007separation}, (e) Empirical Mode Decomposition \cite{mondal2011reduction}, (f) Filtering \cite{nersisson2017heart}, (g) Interpolation \cite{pourazad2006heart}, (h) Modulation Filtering \cite{falk2008modulation}, (i) NMF Clustering 1 \cite{shah2014blind}, (j) NMF Clustering 2 \cite{canadas2017non}, (k) Recursive Least Squares Adaptive Filtering \cite{gnitecki2003recursive}, (l) Singular Spectrum Analysis \cite{ghaderi2011localizing}, (m) Wavelet Transform Based Stationary Non-Stationary Filter \cite{hadjileontiadis1998wavelet}, (n) Wavelet Decomposition and Singular Spectrum Analysis \cite{mondal2017noise}.

Figures~\ref{fig:artificial_heart} and \ref{fig:artificial_lung} show the artificial mixture sound separation results for heart and lung sounds, as detailed in Section~\ref{sec:BSS_eval}. Overall for heart and lung sound separation, both the proposed NMCF (a) and NMF (b) methods significantly outperform existing methods in all situations, with median SDR improvement ranging from 2.7\,dB for respiratory support noise to 11.6\,dB for general noise compared to next best existing method. 

For heart sound separation, in the no noise case, the proposed NMCF (a) and NMF (b) methods outperforms all existing methods except adaptive line enhancement (d), which has a median SDR improvement of 1.5\,dB and 0.8\,dB over proposed methods respectively. However, adaptive line enhancement (d) produces minor temporal distortions in the separated heart sound, resulting in significantly lower SI-SDR values of 18.1\,dB and 17.9\,dB for proposed NMCF (a) and NMF (b) respectively. For the general noise case, proposed NMCF (a) and NMF (b) methods outperform the next best existing method by 4.3\,dB and 4.0\,dB for cry noise and 1.7\,dB and 2.1\,dB for stethoscope movement noise, for median SDR improvement. Similarly for respiratory support noise case, proposed NMCF (a) and NMF (b) methods outperform the next best existing method by 1.8\,dB and 1.4\,dB for bubble CPAP noise and 1.8\,dB and 1.0\,dB for ventilator CPAP noise. 

For lung sound separation, in the noise case, the proposed NMCF (a) and NMF (b) outperformed the next best existing method by 0.9\,dB and 0.8\,dB (not significant p-value=0.055), this increases to 3.1\,dB and 1.4\,dB when for hard to separate situations (lung SNR less than -10\,dB). For the general noise case, proposed NMCF (a) and NMF (b) methods outperform the next best existing method by 8.4\,dB and 6.9\,dB for cry noise and 0.3\,dB and 0.7\,dB for stethoscope movement noise. For high stethoscope movement noise (SNR greater than 0\,dB) median SDR improvement increases to 0.5\,dB and 0.9\,dB.  For respiratory support noise case, proposed NMCF (a) and NMF (b) methods outperform next best existing method by 0.6\,dB and 0.0\,dB (not significant p-value=0.29) for bubble CPAP noise and 1.0\,dB and 0.4\,dB for ventilator CPAP noise. For high respiratory support noise (SNR greater than 0\,dB), median SDR improvement increases to 1\,dB and 0.1\,dB (not significant p-value=0.25) for bubble CPAP noise and 1.2\,dB and 0.9\,dB for ventilator CPAP noise. 

As both proposed NMCF (a) and NMF (b) methods do not produce temporal distortions, SI-SDR results are comparable with SDR. Both methods significantly outperform existing methods in all situations, with median SI-SDR improvement ranging from 5.2\,dB for respiratory support noise to 9.9\,dB for general noise compared to the next best existing method. For specifically heart sound separation or lung sound separation, again, both proposed methods significantly outperformed all existing methods in all situations for SI-SDR improvement. 

For SIR results, both proposed methods significantly outperform existing methods in all situations, with median SIR improvement ranging from 3.8\,dB for respiratory support noise to 8.2\,dB for general noise compared to the next best existing method. For heart sound separation, empirical mode decomposition (e) significantly outperformed both proposed methods in the no noise situation and was comparable for the general noise case. For lung sound separation, the filtering (f) method was comparable to the proposed NMF method for no noise case, but was significantly inferior for hard to separate situations (lung SNR less than -10\,dB). For all other scenarios, both proposed methods significantly outperformed existing methods for heart and lung sound separation.  

For existing methods, the best parameters based on artificial dataset SDR values are as follows: 
\begin{itemize}
    \item Adaptive Fourier Decomposition (c): Decomposition level of 5 (Cry Noise) and 10 (No Noise, Stethoscope Movement Noise, Respiratory Support Noise)
    \item Adaptive Line Enhancement (d): Delay of 1 (No Noise, Respiratory Support Noise), 10 (Stethoscope Movement Noise) and 50 (Cry Noise)
    \item Empirical Mode Decomposition (e): Decomposition using ensemble empirical mode decomposition (All Cases)
    \item Interpolation (g): Remove entire segments containing heart sounds and interpolate (All Cases)
    \item Modulation Filtering (h): bandstop of 3-20\,Hz (All Cases) for heart sounds  and bandpass of 4-20\,Hz (No Noise, Ventilator CPAP and General Noise) and 6-20\,Hz (Bubble CPAP Noise) for lung sounds
    \item Wavelet Transform Based Stationary Non-Stationary Filter (m): Adaptive threshold of 3 (No Noise, Respiratory Support Noise, Cry Noise) and 3.5 (Stethoscope Movement Noise)
\end{itemize}

For proposed NMCF (a) and NMF (b) methods, the best parameters based on artificial dataset SDR values are as follows: 
\begin{itemize}
  \item Time-Frequency Representation = STFT, FFT Size = 1024 samples, Window Size = 512 samples and Hop Size = 256 samples (NMCF All Cases, NMF No Noise, General Noise, Bubble CPAP Noise). Time-Frequency Representation = STFT, FFT Size = 2048 samples, Window Size = 2048 samples and Hop Size = 512 samples (NMF Ventilator CPAP Noise)
  \item Beta loss = 1 (All Cases)
  \item Sparsity penalty = 0.1 (NMCF All Cases, NMF Stethoscope Movement Noise and Respiratory Support Noise), 0 (NMF No Noise and Cry Noise)
  \item NMCF $\lambda_h$ = 0.25 (No Noise), 0 (General Noise, Respiratory Support Noise)
  \item NMCF $\lambda_l$ = 0 (All Cases)
  \item NMCF $\lambda_n$ = 0.25 (Stethoscope Movement Noise, Ventilator CPAP Noise), 0.75 (Cry Noise, Bubble CPAP Noise)
  %\item NMCF $\lambda_n$ for Stethoscope Movement Noise = 0.25
  %\item NMCF $\lambda_n$ for Bubble CPAP Noise = 0.75
  %\item NMCF $\lambda_n$ Ventilator CPAP Noise = 0.25
\end{itemize}

Table~\ref{tab1:regressionresults} shows the real-world chest sound separation results, as detailed in Sections~\ref{sec:HRBR error} and \ref{sec:SQI}. Overall for heart and lung sound separation, all proposed methods significantly outperform existing methods in all situations. Mean signal quality improvement ranged from 0.40-0.63 for no respiratory support case and 0.74-1.12 for respiratory support case compared to the next best existing method. 

For heart sound separation, all proposed methods outperform existing methods. Mean signal quality improvement compared to the next best existing method ranged between 0.10-0.17 for no respiratory support case and 0.22-0.38 for the respiratory support case. All signal quality improvement values were significant except the proposed NMCF (a) with the no respiratory support case (p-value=0.06). 

For lung sound separation, all proposed methods significantly outperformed existing methods except filtering (f). Mean signal quality improvement compared to the filtering method (f) ranged from -0.04-0.14 for the no respiratory support case and 0.00-0.27 for the respiratory support case. For no respiratory case, only the proposed NMF method (b) was significantly better than the filtering method (f). Whereas for the respiratory case, both the proposed NMCF (a) and NMCF with unsupervised components (a un) significantly outperformed the filtering method (f). 

As seen in Table~\ref{tab1:regressionresults} vital sign improvement results were very sporadic. In general, all methods had a median vital sign error improvement of 0, but were positively skewed, as seen in the mean values. Due to the large variation in results, many of the comparisons between methods were not significant. For no respiratory support, both NMCF methods, (a) and (a un), performed best for heart rate estimation (not significant for NMF clustering 2 (j) and filtering (f)). Whereas NMF without supervised components (b un) performed best for breathing rate estimation (not significant for filtering (f) and modulation filtering (h)). For the presence of respiratory support noise, the existing filtering method (f) performed significantly better than all proposed methods for heart rate estimation and wavelet transform based stationary non-stationary filter (m) performed significantly better than proposed NMCF methods, (a) and (a un).

\begin{figure}[tp]
\centering
\includegraphics[scale=0.55,trim={2cm 0.5cm 0.5cm 0.5cm}]{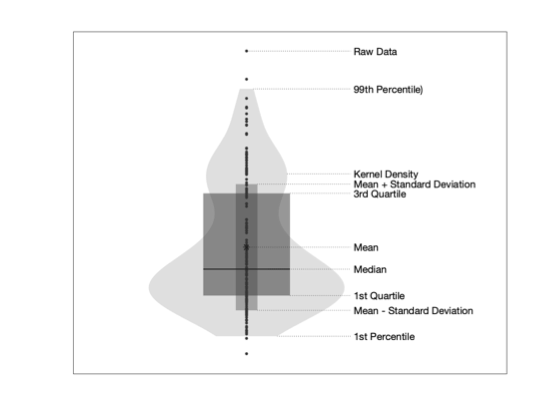}
\caption{Example Violin Plot}
\label{fig:violin_plot}
\end{figure}

%  Letters a to n in the table refer to the following methods: (a) Proposed NMCF, (a un) Proposed NMCF with Unsupervised Components, (b) Proposed NMF, (b un) Proposed NMF with Unsupervised Components, (c) Adaptive Fourier Decomposition \cite{wang2015adaptive}, (d) Adaptive Line Enhancement \cite{tsalaile2007separation}, (e) Empirical Mode Decomposition \cite{mondal2011reduction}, (f) Filtering \cite{nersisson2017heart}, (g) Interpolation \cite{pourazad2006heart}, (h) Modulation Filtering \cite{falk2008modulation}, (i) NMF Cluster 1 \cite{shah2014blind}, (j) NMF Cluster 2 \cite{canadas2017non}, (k) Recursive Least Squares Adaptive Filtering \cite{gnitecki2003recursive}, (l) Singular Spectrum Analysis \cite{ghaderi2011localizing}, (m) Wavelet Transform Based Stationary Non-Stationary Filter \cite{hadjileontiadis1998wavelet}, (n) Wavelet Decomposition and Singular Spectrum Analysis \cite{mondal2017noise}.
\begin{table*}[tb]
    \begin{center}
\caption{Real Noisy Neonatal Chest Sound Results. Mean and standard deviation in brackets is shown for signal quality improvement (SQI) and heart rate (HR) and breathing rate (BR) error improvement as calculated in Sections~\ref{sec:SQI} and \ref{sec:HRBR error}, is shown for heart and lung for recordings with and without respiratory support. Median processing time per 10\,s recording as calculated in Section~\ref{sec:real_time} is also shown. Results in \textbf{bold} refer to best performing sound separation methods. Methods a, a un, b, b un (in blue) are proposed NMCF, NMCF with Unsupervised Components, NMF and NMF with Unsupervised components methods respectively. Methods c to n (in black) are existing methods, as specified in Table~\ref{tab1:existing_methods}.}
    \begin{tabular}{ |M{0.95cm}||M{1.4cm}|M{1.5cm}|M{1.4cm}|M{1.4cm}||M{1.4cm}|M{1.4cm}|M{1.4cm}|M{1.4cm}||M{1.25cm}|}
    \hline
    \multirow{3}{*}{\textbf{Method}} 
    & \multicolumn{4}{c||}{\textbf{Heart}} 
    & \multicolumn{4}{c||}{\textbf{Lung}} 
    & \multirow{3}{*}{\shortstack[c]{\textbf{Processing}\\ \textbf{Time}}}\\
    \cline{2-9}
    & \multicolumn{2}{c|}{\textbf{No Respiratory Support}} 
    & \multicolumn{2}{c||}{\textbf{Respiratory Support}} 
    & \multicolumn{2}{c|}{\textbf{No Respiratory Support}} 
    & \multicolumn{2}{c||}{\textbf{Respiratory Support}} 
    & \\
    \cline{2-9}
    & \textbf{SQI*}
    & \textbf{HR**}
    & \textbf{SQI}
    & \textbf{HR}
    & \textbf{SQI}
    & \textbf{BR***}
    & \textbf{SQI}
    & \textbf{BR}
    & \\
    \hline
    \rowcolor{LightCyan}
    a%NMCF
    & 0.85 (0.94)
    & 3.52 (25.4)
    & 1.28 (0.61)
    & 3.32 (24.5)
    & 0.58 (0.85)
    & -0.19 (10.3)
    & 0.56 (0.69)
    & -0.01 (10.3)
    & 28.2\,s
    \\
    \hline
    \rowcolor{LightCyan}
    a un%NMCF with Unsupervised Noise
    & \textbf{0.92 (1.05)}
    & \textbf{3.64 (25.4)}
    & \textbf{1.34 (0.60)}
    & 1.90 (26.9)
    & 0.62 (0.76)
    & -0.65 (10.9)
    & \textbf{0.65 (0.80)}
    & 0.44 (10.5)
    & 28.3\,s
    \\
    \hline
    \rowcolor{LightCyan}
    b%NMF
    & 0.90 (0.92)
    & 1.39 (25.6)
    & 1.19 (0.59)
    & 1.03 (25.1)
    & \textbf{0.76 (0.93)}
    & \textbf{0.99 (10.6)}
    & 0.42 (0.57)
    & 0.55 (9.1)
    & 275\,ms
    \\
    \hline
    \rowcolor{LightCyan}
    b un%NMF with Unsupervised Noise
    & 0.87 (0.95)
    & 0.72 (25.3)
    & 1.26 (0.57)
    & -0.62 (27.2)
    & 0.67 (0.88)
    & 0.12 (10.8)
    & 0.38 (0.62)  
    & 1.18 (10.8)
    & 342\,ms
    \\
    \hline
    c%Adaptive Fourier Decomposition \cite{wang2015adaptive}
    & 0.48 (1.09)
    & -8.67 (44.9)
    & 0.92 (0.84)
    & 1.18 (32.5)
    & -0.01 (0.69)
    & -1.22 (11.7)
    & -0.11 (0.49) 
    & 1.67 (11.9)
    & 1.30\,ms
    \\
    \hline
    d%Adaptive Line Enhancement \cite{tsalaile2007separation}
    & 0.10 (0.83)
    & -0.17 (23.5)
    & 0.40 (0.74)
    & 2.94 (28.7)
    & -0.01 (0.33)
    & 0.20 (4.0)
    & -0.03 (0.35)
    & 0.06 (3.8)
    & 353\,ms
    \\
    \hline
    e%Empirical Mode Decomposition \cite{mondal2011reduction}
    & 0.18 (0.90)
    & 1.06 (25.6)
    & 0.76 (0.87)
    & 0.98 (28.2)
    & -0.03 (0.68) 
    & -0.53 (7.5)
    & 0.11 (0.54)
    & 0.47 (7.4)
    & 5.53\,s
    \\
    \hline
    f%Filtering \cite{nersisson2017heart}
    & 0.41 (0.90)
    & 1.59 (23.3)
    & 0.18 (0.62)
    & \textbf{7.54 (27.4)}
    & 0.62 (0.90)
    & 0.30 (5.5)
    & 0.38 (0.64)
    & 0.02 (6.4)
    & 2.90\,ms
    \\ 
    \hline
    g%Interpolation \cite{pourazad2006heart}
    & 0.24 (0.98)
    & 0.57 (27.32)
    & 0.73 (0.76)
    & 5.21 (34.2)
    & -0.18 (0.66)
    & -1.17 (5.2)
    & -0.09 (0.53)
    & 0.84 (19.2)
    & 310\,ms
    \\
    \hline
    h%Modulation Filtering \cite{falk2008modulation}
    & 0.12 (0.81)
    & -2.85 (25.7)
    & 0.47 (0.65)
    & 0.77 (33.7)
    & 0.05 (0.65)
    & 0.33 (8.5)
    & -0.06 (0.44)
    & -0.40 (9.3)
    & 67.5\,ms
    \\
    \hline
    i%NMF Cluster 1 \cite{shah2014blind}
    & 0.55 (0.86)
    & -1.88 (30.7)
    & 1.02 (0.57)
    & -2.49 (32.7)
    & 0.06 (0.60) 
    & 0.21 (8.3)
    & 0.00 (0.48)
    & 0.38 (7.09)
    & 876\,ms
    \\
    \hline
    j%NMF Cluster 2 \cite{canadas2017non}
    & 0.21 (0.74)
    & 2.46 (23.6)
    & 0.64 (0.67)
    & 1.77 (28.7)
    & 0.37 (0.61) 
    & -1.02 (9.6)
    & 0.15 (0.48)
    & -0.97 (9.5)
    &  5.26\,s
    \\
    \hline
    k%Recursive Least Squares Adaptive Filtering \cite{gnitecki2003recursive}
    & 0.25 (1.00)
    & 0.48 (24.3)
    & 0.70 (0.71) 
    & 3.01 (32.9)
    & 0.01 (0.55)
    & -0.07 (6.3)
    & 0.12 (0.53)
    & 0.37 (6.3)
    & 6.51\,s
    \\
    \hline
    l%Singular Spectrum Analysis \cite{ghaderi2011localizing}
    & 0.75 (0.94)
    & -11.03 (42.2)
    & 0.75 (0.78)
    & 2.48 *41.0)
    & -0.20 (0.71)
    & -4.2 (14.7)
    & 0.09 (0.50)
    & 0.07 (8.5)
    &  402\,ms
    \\
    \hline
    m%Wavelet Transform Based Stationary Non-Stationary Filter \cite{hadjileontiadis1998wavelet}
    & 0.20 (0.90)
    & -5.14 (25.2)
    & 0.88 (0.66)
    & -5.11 (35.3)
    & -0.39 (0.62)
    & -0.92 (15.4)
    & -0.17 (0.60)
    & \textbf{3.46 (15.1)}
    &  36.9\,s
    \\
    \hline
    n%Wavelet Decomposition and Singular Spectrum Analysis \cite{mondal2017noise}
    & 0.38 (1.01)
    & -6.92 (37.8)
    & 0.96 (0.92)
    & 4.13 (33.1)
    & NA
    & NA 
    & NA
    & NA
    &  531\,ms
    \\
    \hline
    \multicolumn{10}{p{0.8\textwidth}}{*SQI= Signal Quality Improvement (Section~\ref{sec:SQI}), **HR= Heart Rate Error Improvement Improvement (Section~\ref{sec:HRBR error}), ***BR= Breathing Rate Error Improvement (Section~\ref{sec:HRBR error})}\\    
    \end{tabular}
    \label{tab1:regressionresults}
\end{center}    
\end{table*}

\section{Discussion}
\label{sec:discussion}

% Comparision between NMCF, NMF 
Overall, both proposed NMCF and NMF methods performed well, producing comparable or superior  results to existing methods. In the artificial dataset, NMCF had a median SDR improvement of 0.00, 1.71, -0.55, 1.17 and 0.92\,dB for no noise, cry noise, stethoscope movement noise, bubble CPAP noise and ventilator CPAP noise, respectively in comparison to NMF. Whereas in the real-world dataset, NMF outperformed NMCF in the no respiratory support case, and NMCF outperformed NMF in the respiratory support case. As seen in the results, in the presence of hard-to-separate noise such as respiratory support noise, co-factorisation in the NMCF method aids in the better adaption for effective sound separation, as opposed to the NMF method. Whereas, in the general case of just heart-lung sound separation, the NMF method is sufficient for successful sound separation. As the proposed NMF method can be implemented in real-time, it would be recommended to utilise this method initially. If the signal quality is still poor after the implementation of NMF due to hard-to-separate noise, then the proposed NMCF method should be utilised afterwards. NMCF allows for better adaptation of the basis vectors, as opposed to fixed basis vectors in the NMF method, enabling better ability to segregate noise. 

% Vital Sign error
Vital sign error results were less conclusive and promising for the purposed methods. With regards to median vital sign error improvement being 0, in general, the vital sign estimation methods are designed to be robust to an extent to noise. In particular, the heart rate estimation method was designed specifically to deal with noisy heart sounds \cite{springer2015logistic}. Instead, the sound separation and denoising methods mainly only improved results when there was a major vital sign estimation error in the raw recordings, explaining the large standard deviation and positive mean seen in the results. For sound separation in the presence of respiratory support noise, the proposed NMF and NMCF methods were inferior to existing methods. One possible explanation for this is because after separation, with respiratory support noise partially removed, there were cases where heart and lung sounds were not present, resulting in an underestimation in vital signs. Whereas, for other methods that do not explicitly remove the respiratory support sounds, these sounds were miscounted as heartbeats/breathing periods, avoiding the underestimation in vital signs. These conclusions are further supported by existing methods that poorly estimated vital signs in no respiratory support case showing improvement in the more difficult respiratory support case. In particular for lung sounds, all methods struggled to successfully obtain clean lung sounds for the respiratory support case, suggesting all breathing rate error improvements could be inaccurate. Future work is required to accurately determine if these removed respiratory support sounds are also unintentionally removing these heartbeats/breathing periods, or if there is a subset of recordings in which heart and lung sounds cannot be recovered. 

% NMCF un, NMF un
As seen in Table~\ref{tab1:regressionresults}, the introduction of the unsupervised noise component led to an improvement in signal quality results and a decrease in signal quality for the NMF method. For the NMCF, the benefit of the unsupervised noise component can be explained by the fact that it prevents co-factorisation and learning of the basis vectors and temporal activation to be heavily affected by noise that was not present in the reference sound set. Instead, other noise sources are allocated to the unsupervised component as they are not related to any of the reference sounds. For NMF, the basis matrix has already been pre-computed and fixed for the sound separation phase. Hence, this method is not as affected by the presence of other sources of noise, especially if they have fairly distinct frequency components. Instead, relevant heart and lung sound components can be misplaced into the unsupervised components, due to the unsupervised nature of calculating these basis vectors, resulting in a decrease in performance. This is a key limitation of the introduction of unsupervised components, as whilst it can remove unknown noise sources, it can also remove some of the heart and lung sound components. 

 % Improvements and Future work on NMF and NMCF
 Whilst our proposed sound separation methods worked better or comparably to existing methods for respiratory support noise, there is still a large room for improvement. In both the artificial dataset and real-world recordings, respiratory support noise was still present in the separated sounds. This can be explained by the fact that respiratory support noise has a large frequency overlap with the desired heart and lung sounds. This large frequency overlap results in the frequency basis vectors being quite similar for the respiratory support noise and heart and lung sounds. Some possible ways to improve sound separation in the presence of respiratory support noise is to have additional reference sounds and add a temporal constrain or decomposition. Currently, only 5 and 8 reference sounds were provided for bubble and ventilator CPAP respectively, which may not provide the entire variety of different flow rates and sounds that can occur. With regards to temporal constraint or decomposition, respiratory support noise tends to be present for the entire recordings, whereas heart and lung sounds occur on and off. Therefore, there is a potential for reassignment of sub-components after NMF or NMCF based on temporal activation, similar to past works  \cite{tsai2020blind,lin2017improving}. Additionally, instead of reassignment post decomposition, there is the possibility of adding a temporal constraint or cost function, which would enable the decomposition of the components based on temporal activation. However, future work would be required to develop this temporal decomposition method.

% Computational time
 With regards to computational cost, the proposed NMCF method is not suitable for real-time processing, whereas the proposed NMF method is. Given, for the most part, the proposed NMF method produces superior results to all existing methods and comparable results to the proposed NMCF method, this may be sufficient. There are a couple of ways to improve the computation speed of the proposed NMCF method. Firstly, as the optimal heart and lung sound co-factorisation weights were zero (i.e., no co-factorisation with mixture chest sound recording), a fusion of the NMF and NMCF could be employed. For these zero co-factorisation weights, instead of calculating the associated basis vectors during the separation of the mixture chest sound, instead, similar to NMF, the basis matrices of heart and lung sounds can be pre-computed and fixed. Accordingly, the computation time would be reduced by up to 36\% potentially. Secondly, the pre-computed base matrices could be used as the initialisation of the basis matrices for the proposed NMCF method. This initialisation would place the basis matrices closer to the optimal result, meaning fewer iteration of the update algorithm would be required. Thirdly, there is still room for optimisation of the number of reference examples, the number of bases for each sound component, FFT size and hop size, all of which could reduce the size of the basis and activation matrices, overall reducing the computation time. Finally, the multiplicative update algorithm could be optimised, for instance using C code MEX functions, as opposed to MATLAB. 

 By and large, all these conclusions on computational time with regards to real-time processing is specific to the laptop parameters used in this particular study. It would be expected similar computational times for high-end laptops, general desktops, or a general device connected to cloud computing, however, computational times would be expected to increase if these sound separation methods were used on a smartphone with its onboard processing power.

% Artifical dataset benefit and limitations
An artificial dataset was proposed to quantitatively assess the effectiveness of sound separation with heart, lung and various noise sounds. However, there are several limitations with this dataset. 
Firstly, to obtain clean reference sounds, bandpass filtering of 50-250\,Hz and 200-1000\,Hz was applied to heart and lung sounds. Similarly, for cry noise, highpass filtering of 300\,Hz was applied. Whilst these frequency bands contain the majority of the heart, lung and cry sounds, there can still be prominent sounds outside these frequency bands. Therefore, to more accurately replicate the larger frequency overlap between these sounds and assess the sound separation methods, pure and unfiltered reference sounds would be required. 
However, this is quite difficult to obtain with newborns. For example, clean reference heart sounds are typically acquired from adults by asking them to hold their breath, which is not feasible with neonates. However neonates, in particular preterm neonates, have sporadic breathing patterns of fast periodic breathing followed by a period of no breathing. Therefore, it may be possible to obtain short segments of clean heart sounds. Obtaining clean reference lung sounds is more difficult as heart sound noise will always be present. Future collection of chest sounds with digital stethoscope placement on the chest but further away from the heart or on the back may minimise heart sounds and enable clearer lung sounds. Another option to obtain clean reference sounds with a full frequency range is to artificially generate the sounds with the desired properties. Past works on modelling adult heart sounds have been done and could be transferable for neonates. Similarly, a model of neonatal lung sounds and various noise sources could be created. 

The second limitation with the artificial dataset is how cry noise was mixed with lung sounds. As inspiration and expiration cannot occur simultaneously as a newborn is crying, this should be taken into account. While for stethoscope disconnect noise, this was resolved by zeroing out the section of the heart-lung sound mixture that it was going to be inserted, this is not appropriate for cry noise. For a more appropriate lung sound and cry noise mixture, the periods and order of inspiration and expiration should be taken into account and inserted reasonably. However, as lung sounds and breathing rate change while a baby is crying, it may not even be useful or necessary to try and accurately represent this mixture. 

The third limitation with the artificial dataset is the relatively small evaluation dataset, in particular for respiratory support noise and heart and lung sounds. For respiratory support noise, future collections of pure respiration support noise obtained by placing the stethoscope on the breathing tube will be acquired. Whereas the limitation with heart and lung sounds is not the number of chest sound recordings available, but the difficulty in obtaining clean sounds. Again, one potential solution to this is to artificially create heart and lung sounds. 

% Real SQI benefits and limitations
Furthermore, signal quality estimation of real chest sounds was proposed to quantitatively assess the effectiveness of sound separation on real-world data. This is an essential assessment as it overcomes many of the limitations with artificial datasets and assesses the sound separation methods on the actual data it would be used for. However, there are also several limitations with the proposed signal quality estimation method. Firstly, while the method attempts to objectively assess heart and lung sound quality, it is not 100\% accurate \cite{grooby2020neonatal, grooby2021real}. Therefore, individual comparison of methods with a single recording may not be appropriate due to misclassification errors with the signal quality estimation model. Whereas, overall trends on signal quality improvement and comparison of methods is more appropriate. Secondly, the method assesses the overall quality of the heart and lung sounds. This overall assessment does not directly take into account if significant portions of the heart or lung sound components are also being removed to obtain the "noise-free" sounds. These limitations can also partially explain why the artificial dataset vs real-world dataset sound separation methods differs.

\section{Conclusion}
\label{sec:conclusion}
This paper has reviewed, adapted and tested a wide variety of existing methods for neonatal chest sound separation for clean heart and lung sounds. Additionally, two methods were  proposed, namely, NMF and NMCF with reference datasets of heart, lung and noise sounds. The evaluation results show, our proposed methods outperformed existing methods in both the artificial and real-world datasets with regards to SDR, SIR and SI-SDR, and signal quality, respectively. However, further work is still required for sound separation of respiratory support noise, as can be seen in the inferior results compared to existing methods for vital sign estimation. Possible directions are obtaining more clean samples either in the real-world setting or artificially and looking into deep learning methods. 

\section*{Acknowledgment}

E. Grooby thanks and acknowledges the support from the Monash Newborn team at Monash Children's Hospital, Australia for data collection and Jinyuan He for the analysis of the newborn data in previous works. 

\bibliographystyle{IEEEtran}
\bibliography{IEEEabrv,references}
\end{document}